\documentclass[12pt,fleqn]{article}

\usepackage{graphicx}
\usepackage{dcolumn}
\usepackage{bm}
\usepackage{placeins}
\usepackage{amsmath}
\usepackage{authblk}

\begin{document}

\title{Thermodynamic curvature of the binary van der Waals fluid}

\author{George Ruppeiner\footnote{Division of Natural Sciences, New College of Florida, 5800 Bay Shore Road, Sarasota, FL 34243, USA (ruppeiner@ncf.edu)}\,\, and Alex Seftas\footnote{Department of Physics, Applied Physics, and Astronomy, Rensselaer Polytechnic Institute, 110 Eighth Street, Troy, NY, 12180-3590, USA}}

\date{\today}

\maketitle

\begin{abstract}

The thermodynamic Ricci curvature scalar $R$ has been applied in a number of contexts, mostly for systems characterized by 2D thermodynamic geometries. Calculations of $R$ in thermodynamic geometries of dimension three or greater have been very few, especially in the fluid regime. In this paper, we calculate $R$ for two examples involving binary fluid mixtures: a binary mixture of a van der Waals (vdW) fluid with only repulsive interactions, and a binary vdW mixture with attractive interactions added. In both these examples, we evaluate $R$ for full 3D thermodynamic geometries. Our finding is that basic physical patterns found for $R$ in the pure fluid are reproduced to a large extent for the binary fluid.

\end{abstract}

\section{Introduction}

The thermodynamic Ricci curvature scalar $R$ has yielded a number of interesting results in the study of fluids \cite{Ruppeiner1995}. However, to this point, the great majority of the calculations of $R$ have been made for pure fluids. Much less examined has been $R$ for binary fluids. Pure fluids offer many research topics in a relatively simple thermodynamic geometric scenario. One element of this simplicity is that pure fluids may be represented by a two-dimensional (2D) thermodynamic Riemannian geometry, where just the scalar $R$ gives the full curvature picture.

\par
The dimension of the thermodynamic phase space grows by one for each added fluid component, and, as the dimension grows, the curvature rapidly becomes more complicated. For example, the binary fluid corresponds to a three-dimensional (3D) phase space, where there are six independent components of the full Riemannian curvature tensor \cite{Carroll2004}. Which of these components do we focus on for physical interpretation? In this paper, our results suggest that the curvature scalar $R$ remains the fundamental physical quantity.\footnote{We offer no rigorous proof, but we do note that classical general relativity, based on four-dimensional Riemannian geometry, may be expressed as a variational principle based on the Hilbert action, with $R$ as the Lagrangian \cite{Carroll2004}. This variational principle is certainly not in play in this paper, but we do hope that this significance for $R$ translates to the thermodynamic scenario as well.}
 
\par
We calculate $R$ for two thermodynamic examples represented by a 3D thermodynamic geometry: 1) a binary van der Waals (vdW) fluid with just repulsive interactions, and 2) a binary vdW fluid with attractive interactions added. To add conceptual structure to our presentation, we discuss to what extent the 3D results for $R$ follow the same physical principles as the 2D ones. As we show, there is a great deal of correspondence.\footnote{Scenarios with a thermodynamic geometric dimension greater than two have also been considered for black hole thermodynamics; see Sahay for a brief review \cite{Sahay2017a}. But this theme is beyond the scope of this paper.}

\par
A few cases have been worked out for $R$ for the full 3D thermodynamic geometry. Ruppeiner and Davis \cite{Ruppeiner1990} worked out $R$ for the multicomponent ideal gas, with an arbitrary number of components. Kaviani and Dalafi-Rezaie \cite{Kaviani1999} worked out $R$ for the paramagnetic ideal gas, where, in addition to the temperature and the density, there is an external organizing magnetic field. Erdem \cite{Erdem2018} worked out $R$ for the antiferromagnetic Ising model with a temperature and two ordering fields, with special attention to the critical phenomena near the Neel point.

\par
One may also approach 3D thermodynamic geometries by working out $R$ over 2D slices of the full 3D thermodynamic geometry. This has physical relevance if, for example, one of the independent variables is irrelevant to a phase transition. Ginoza \cite{Ginoza1993} considered the binary fluid mixture in generality, and calculated $R$ 2D along surfaces of constant $T$. Jaramillo-Gutiérrez \cite{Arenas2020}, et al., calculated a constrained $R$ 2D by fixing the sum of the mole numbers. These authors made several comments about how their constrained $R$ relates to that of the pure fluid. Generally, the calculation is simplified in 2D, but, in an age of powerful mathematics software, considerations of simplicity should maybe no longer be so much of a driving force.

\section{Expectations for 3D outcomes for $R$}

\par
For guidance as to what 3D thermodynamic geometry might tell us, let us pose some questions motivated by findings in the pure fluid \cite{Ruppeiner2010,Ruppeiner2012a, Ruppeiner2012b, May2013}: a) In the ideal gas limit, is $|R|$ either zero or small? ``Small’’ means $|R|\rho<1$, with $\rho$ the density. In this event, the volume measured by $|R|$ is less than the average volume occupied by a single particle, and we are under the low $|R|$ limit \cite{Ruppeiner2012a, Ruppeiner2012b}. Thermodynamics is challenged at such a size scale, and a precise physical interpretation for $R$ is harder to come by (though researchers do try if the opportunity arises). b) As a critical point is approached, does $|R|$ diverge in proportion to the correlation volume $\xi^3$? c) Near a critical point, is the sign of $R$ negative? A negative sign is consistent with effectively attractive interactions. d) Are the values of $R$ in two coexisting phases equal to each other near a critical point? e) Are there interesting regimes of positive $R$? A positive sign is consistent with effectively repulsive interactions. f) Does the binary repulsive fluid have instances of anomalous negative $R$, such as is present for the hard-sphere pure fluid \cite{Branka2018}?

\par
We find considerable physical correspondence between $R$ for the pure fluid and for the binary fluid.

\section{Thermodynamic geometry of a binary fluid}

For a binary fluid, the Helmholtz free energy per volume may be written in terms of its appropriate coordinates \cite{Ruppeiner1990, Callen1985},
\begin{equation}
    f=f(T, \rho_{1}, \rho_{2}),
    \label{eqn:Helmholtz free energy}
\end{equation}

\noindent where $T$ is the temperature, and $\rho_{1}=n_1/V$ and $\rho_{2}=n_2/V$ are the molar densities of the two components $1$ and $2$. Here $n_1$ and $n_2$ are the mole numbers of the components, and $V$ is the volume.

\par
\begin{figure}
\centering
\includegraphics[width=4in]{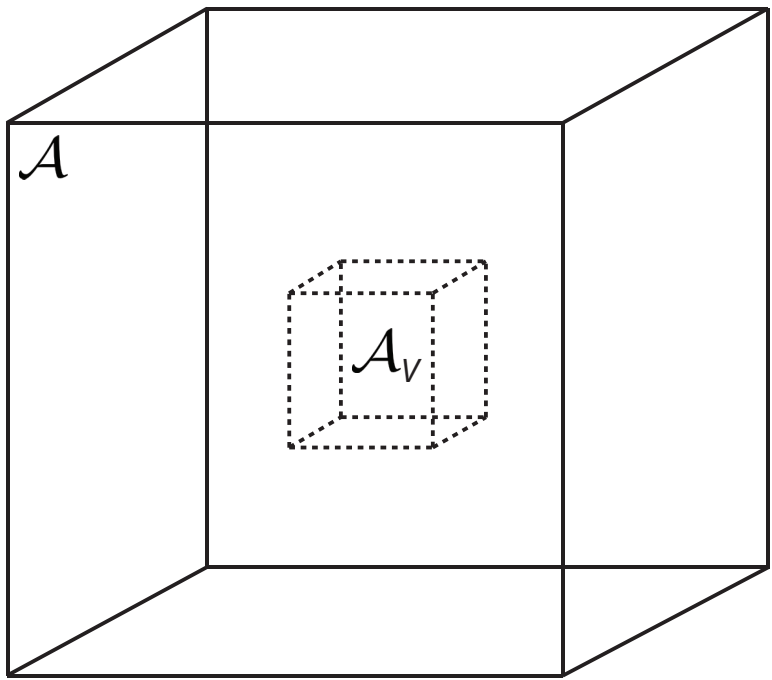}
\caption{A standard structure in thermodynamic fluctuation theory, a single open subsystem $\mathcal{A}_V$, with fixed volume $V$, of an infinite closed environment $\mathcal{A}$.}
\label{Figure1}
\end{figure}

The thermodynamic entropy metric originates from the thermodynamic fluctuation theory \cite{Ruppeiner1995, Landau1980}. Consider a finite open subsystem $\mathcal{A}_V$, with fixed volume $V$, of an infinite closed thermodynamic fluid system $\mathcal{A}$; see Figure \ref{Figure1}. The thermodynamic state $(T, \rho_{1}, \rho_{2})$ of $\mathcal{A}_V$, fluctuates in time about the state of $\mathcal{A}$. The Gaussian fluctuation probability density is proportional to
\begin{equation}\exp\left[-\frac{V}{2}(\Delta\ell)^2\right],\end{equation}

\noindent where the entropy metric \cite{Ruppeiner1995}
\begin{equation}
    (\Delta \ell)^2 = \frac{1}{k_B T}\left(\frac{\partial s}{\partial T}\right)(\Delta T)^2 + \frac{1}{k_B T}\sum_{i,j = 1}^{2}\left(\frac{\partial\mu_i}{\partial\rho_j}\right)\Delta\rho_i\Delta\rho_j.
    \label{eqn:Riemannian metric}
\end{equation}

\noindent Here, the entropy per volume of $\mathcal{A}_V$ is
\begin{equation}
    s = -\frac{\partial f}{\partial T},
    \label{eqn:entropy Maxwell relation}
\end{equation}

\noindent and the chemical potentials of the two fluid components are 
\begin{equation}
    \mu_i = \frac{\partial f}{\partial \rho_i}.
    \label{eqn:chemical potential Maxwell relation}
\end{equation}

\noindent $\Delta T$, $\Delta \rho_1$, and $\Delta \rho_2$ are the differences in the temperature and density coordinates of $\mathcal{A}_V$ and $\mathcal{A}$. Also, $k_B$ is Boltzmann's constant. The thermodynamic metric is employed here in fluctuation theory. But it has also been used in finite time thermodynamics \cite{Andresen1984} as a measure of dissipation.

\par
Let us emphasize a point not always appreciated in the metric geometry of thermodynamic fluctuations. A major project in this geometry is to calculate the thermodynamic curvature $R$. We want $R$ to tell us something about the intrinsic properties of the material comprising the system $\mathcal{A}$. This goal naturally requires the use of an open subsystem $\mathcal{A}_V$, so as to leave particles free to move in or out of $\mathcal{A}_V$, unimpeded by any surrounding artificial wall or membrane. Such a physical constraint would change the value of $R$ in a way involving more than just the properties of the particles; this should be avoided.

\par
As emphasized by Callen \cite{Callen1985}, fluid thermodynamics requires us to set one of the thermodynamic variables aside as the fixed subsystem scale. For an open system, this scale is the volume $V$. We recommend that authors always work with open subsystems for calculating $R$.

\par
In the coordinates $(T, \rho_1, \rho_2)$, the metric tensor \textbf{g} is composed of five nonzero elements, which can be read off from Eq. (\ref{eqn:Riemannian metric}):
\begin{equation}
\textbf{g}=\frac{1}{k_B T}\displaystyle
    \left({\begin{array}{ccc}
   -\frac{\partial ^2 f}{\partial T^2} & 0 & 0\\
    0 & \frac{\partial ^2 f}{\partial \rho_1^2} & \frac{\partial ^2 f}{\partial \rho_1 \partial \rho_2}\\
    0 & \frac{\partial ^2 f}{\partial \rho_2 \partial \rho_1} & \frac{\partial ^2 f}{\partial \rho_2^2}
    \end{array}}\right).
    \label{eqn:metricarray}
\end{equation}

\noindent The fourth-rank curvature tensor has elements \cite{Carroll2004}
\begin{equation}
    R^\alpha_{\beta\gamma\delta} = \Gamma^\alpha_{\beta\gamma,\delta} -  \Gamma^\alpha_{\beta\delta,\gamma} + \Gamma^\mu_{\beta\gamma}\Gamma^\alpha_{\delta\mu} - \Gamma^\mu_{\beta\delta}\Gamma^\alpha_{\gamma\mu},
\end{equation}

\noindent where the Greek indices range from $1,2,3$, and denote the coordinates $(T,\rho_1,\rho_2)$, respectively. The Christoffel symbols are defined as
\begin{equation}
    \Gamma^\alpha_{\beta\gamma} = \frac{1}{2}[g^{\alpha\mu}(g_{\mu\gamma,\beta} + g_{\mu\beta,\gamma} - g_{\beta\gamma,\mu})],
\end{equation}

\noindent where $g^{\alpha\beta}$ denotes the components of the inverse of the metric {\bf g} with components $g_{\alpha\beta}$. We use the comma notation (e.g. $,\beta)$ to denote the partial derivative with respect to some specific coordinate. Repeated indices are summed over. The second-rank Ricci tensor is given by
\begin{equation}
    R_{\alpha\beta} =  R^\mu_{\alpha\mu\beta},
\end{equation}

\noindent and the Ricci curvature scalar is
\begin{equation}
    R = g^{\mu\nu}R_{\mu\nu}.
    \label{R}
\end{equation}

\par
It is straightforward to show that $R$ has units of volume per particle, the same units as for the pure fluid. These units alone label $R$ as a measure of some sort of size scale within the system. The definition of the Riemannian curvature tensor is ambiguous with respect to a sign, and we use the sign convention of Weinberg \cite{Weinberg1972}, where the two spheres have negative $R$.

\section{Thermodynamic stability}

\par
Fluctuations must satisfy thermodynamic stability, requiring maximum entropy of $\mathcal{A}$ in equilibrium. This is obtained if the line element in Eq. (\ref{eqn:Riemannian metric}) is positive definite for all fluctuations. A necessary and sufficient condition for thermodynamic stability is that the metric coefficients in Eq. (\ref{eqn:metricarray}) satisfy the three following conditions \cite{Ruppeiner1995,Eves1966}

\begin{equation} g_{11}>0, \label{Stability1}\end{equation}

\begin{equation}
\begin{array}{|cc|}
    g_{11} & g_{12}\\
    g_{21} & g_{22}
\end{array}>0,
\label{Stability2}
\end{equation}

\noindent and

\begin{equation}
\begin{array}{|ccc|}
    g_{11} & g_{12} & g_{13}\\
    g_{21} & g_{22} & g_{23}\\
    g_{31} & g_{32} & g_{33}
\end{array}>0.
\label{Stability3}
\end{equation}

\noindent We tested frequently for stability.

\section{Helmholtz free energy for binary vdW}

We take the Helmholtz free energy per mole for binary vdW from Konynenburg and Scott \cite{Scott1980}:

\begin{equation} \begin{array}{lr} \displaystyle A_m = e(T)-\mathcal{R}\,T \log\left(\frac{V_m-b}{V_{m0}}\right)\\ \displaystyle
\hspace{3 cm} + \, \mathcal{R}\,T \left[(1-x) \log(1-x)+x \log (x)\right] -\frac{a}{V_m}, \end{array} \label{MolarA}\end{equation}

\noindent where
\begin{equation} e(T)=-\frac{3}{2}\,\mathcal{R}\,T \ln\left(\frac{T}{T_0}\right) + \epsilon \label{IdealGas}\end{equation}

\noindent is the purely thermal part of the ideal gas contribution, $\mathcal{R}$ is the universal gas constant, $V_m=V/(n_1+n_2)$ is the molar volume, $V_{m0}$, $T_0$, and $\epsilon$ are constants that do not appear in $R$, and $x$ is the concentration
\begin{equation}x=\frac{n_2}{n_1+n_2}.\label{Concentration}\end{equation}

\noindent The quantities $a$ and $b$ are functions of $x$ reflecting the purely attractive and repulsive parts of the interparticle interactions, respectively:
\begin{equation} a(x)=(1-x)^2 a_{11} + 2x(1-x)\,a_{12} + x^2 a_{22},\end{equation}

\noindent and
\begin{equation}b(x)=(1-x)\,b_{11}+x\, b_{22},\end{equation} 

\noindent with $a_{11}$, $a_{12}$, $a_{22}$, $b_{11}$, and $b_{22}$ as the five independent vdW coefficients.

\par
The full (extensive) Helmholtz free energy is
\begin{equation}A(T,n_1,n_2,V)=(n_1+n_2)A_m, \label{Helmholtz}\end{equation}

\noindent and the Helmholtz free energy per volume is
\begin{equation}f(T,\rho_1,\rho_2)=\frac{A(T,n_1,n_2,V)}{V}.\label{fpervol}\end{equation}

\par
For the pure fluid, vdW simplifies by using scaled units \cite{Landau1980}. For binary vdW, scaling cannot be done in a natural way. One possibility is to scale in terms of the van der Waals coefficients of one of the fluid components \cite{Scott1980}, but this seems somewhat artificial. In this paper, we avoid scaled units, and simply adopt real units when needed.

\section{Repulsive binary van der Waals}
\par
First consider the case with only repulsive interactions, with zero $a$ coefficients, $b_{11}= b_{1}$, and $b_{22}=b_{2}$. By Eq. (\ref{fpervol}), we have

\begin{equation} \begin{array}{lr} \displaystyle f(T,\rho_1,\rho_2) = -k_{B} T (\rho_1+\rho_2) \log \left[\frac{1-b_1 \rho_1- b_2 \rho_2}{V_{m0}}\right]

\\ \hspace{3 cm} \displaystyle +k_{B} T\rho_1 \log\left(\rho_1 \right)+ k_{B} T\rho_2 \log
   \left(\rho_2\right)
   
\\ \hspace{3.4 cm}\displaystyle -\frac{3}{2} k_{B} T (\rho_1+\rho_2) \log \left(\frac{T}{T_0}\right)+\epsilon (\rho_1+\rho_2).
 \end{array} \label{fHard}\end{equation}

\noindent where we send $\mathcal{R}\to k_B $ to convert $\rho_1$ and $\rho_2$ to units of particles per volume.

\par
By using Eq. (\ref{fHard}), and the process described in section 3, we find:
\begin{equation}
\begin{split}
     R = \frac{1}{2 \left(\rho _1+\rho _2\right)}\left[1-2 b_2 \rho_1-2 b_1 \rho_2-4 b_1 \rho_1-4 b_2 \rho_2+5 b_1^2 \rho_1 \rho_2\right.\\\left.+5 b_2^2 \rho_1 \rho_2+5 b_1^2 \rho_1^2+5 b_2^2 \rho_2^2\right].
   \label{Riemannian Curvature Scalar Hard Sphere}
\end{split}
\end{equation}

\noindent Note, $R$ for repulsive vdW has no dependence on $T$, nor does it depend on the constants in Eq. (\ref{fHard}), other than the $b$'s.
\par
Because $R$ has units of volume per particle, $R \times (\rho_1 + \rho_2)$ is dimensionless. Setting $b_1 = b_2 = b$ yields
\begin{equation}R\times(\rho_1+\rho_2)=\frac{1}{2} \left[1-b\left(\rho_1 +\rho_2\right)\right]\left[1 - 5 b \left(\rho_1 +  \rho_2\right)\right].\label{simpleR} \end{equation}

\begin{figure}[!ht]
\centering
\includegraphics[width=4in]{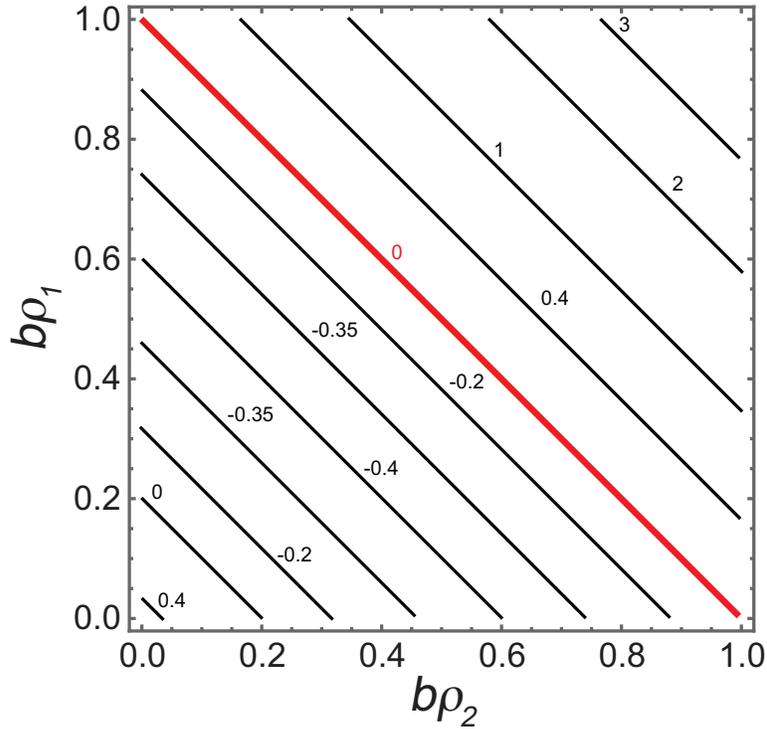}
\caption{The contour graph of $R \times (\rho_1 + \rho_2)$. On approaching the bold red line from below, the pressure $P\to +\infty$. Above the line $P$ is negative. The physical regime is below the line, where $R$ has both positive and negative values, with a minimum of $R \times (\rho_1 + \rho_2)=-0.4$. In the physical regime, values of $R \times (\rho_1 + \rho_2)$ are all less than the low $|R|$ limit. The value at the origin is $1/2$.}
\label{Figure2}
\end{figure}

\noindent If $b\to 0$, this expression agrees with Ruppeiner and Davis \cite{Ruppeiner1990} for the binary ideal gas with the thermal contribution in Eq (\ref{IdealGas}). $R \times (\rho_1 + \rho_2)$ is shown as a contour graph in Figure \ref{Figure2}.

\par
With only the $b$ coefficients nonzero, we would naively expect $R>0$, consistent with repulsive interactions. But cases with $R<0$ clearly occur, mirroring the situation in the pure fluid where, for example, the gas of hard-spheres has negative $R$ \cite{Branka2018}. Such anomalous results might be dismissed as aberrations since we are below the low $|R|$ limit. However, it is interesting that the negative sign persists from the pure fluid into the binary fluid. This consistency may well indicate the need for a more nuanced interpretation for the sign of $R$.

\section{Attractive binary van der Waals}

Now turn on the attractive interactions in binary vdW. In this scenario, Konynenburg and Scott \cite{Scott1980} classified nine distinct possibilities, depending on the values of the vdW $a$ and $b$ coefficients. We consider here only an instance of their Type I possibility, with a single curve of critical points continuously connecting the critical points of the pure fluid components $x=0$ and $x=1$. Our example has vdW coefficients $a_{11}=0.002$, $a_{22}=0.005$, $a_{12}=0.004$, and $b_{11}=b_{22}=0.00002$.\footnote{The units of the $a$'s are Joules $\mbox{meters}^3/\mbox{mole}^2$, and the units of the $b$'s are $\mbox{meters}^3/\mbox{mole}$. These vdW coefficients produce critical points very roughly in the zone of normal fluid Helium.}

\par
\begin{figure}
\centering
\includegraphics[width=4in]{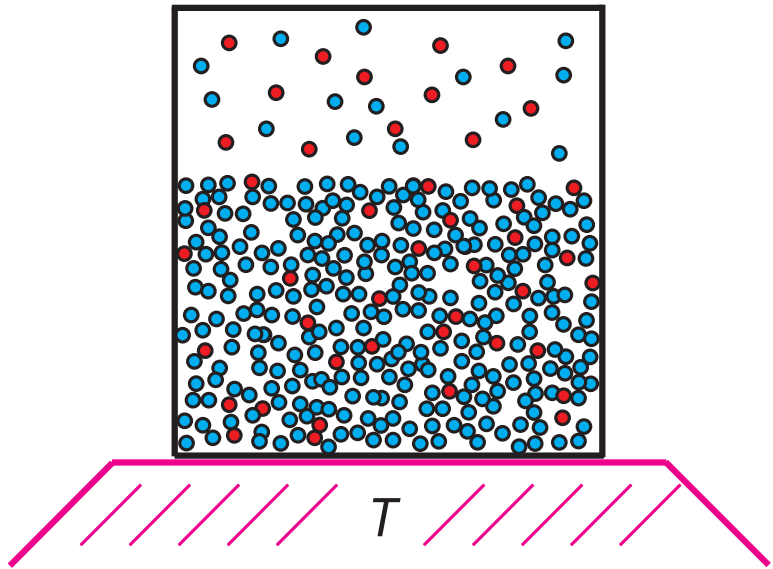}
\caption{The binary fluid inside a closed container fixing the number of particles and the volume. As the temperature is lowered, the pressure decreases, and the fluid generally breaks into two phases at some temperature. The phases have different $x$'s.}
\label{Figure3}
\end{figure}

To structure the discussion, note that if $T$ is high, the attraction between the particles has little effect, and we expect no phase transitions. Now consider lowering $T$ slowly, with the particle numbers and the volume held fixed. We might move along such an isochore in a laboratory $PVT$ experiment; see Figure \ref{Figure3}. As $T$ decreases, attractive interatomic interactions become more effective, and the binary fluid could eventually break into two phases.

\par
These phase transitions have associated second-order curves of critical points \cite{Scott1980}. For Type 1, every $x$ has a single critical point with critical coordinates $T=T_c$, $P=P_c$, and $V_m=V_{mc}$. In their Appendix A, Konynenburg and Scott \cite{Scott1980} describe an explicit procedure $x\to(T_c,P_c,V_{mc})$ for locating these critical points.

\par
\begin{figure}
\begin{minipage}[b]{0.5\linewidth}
\includegraphics[width=2.7in]{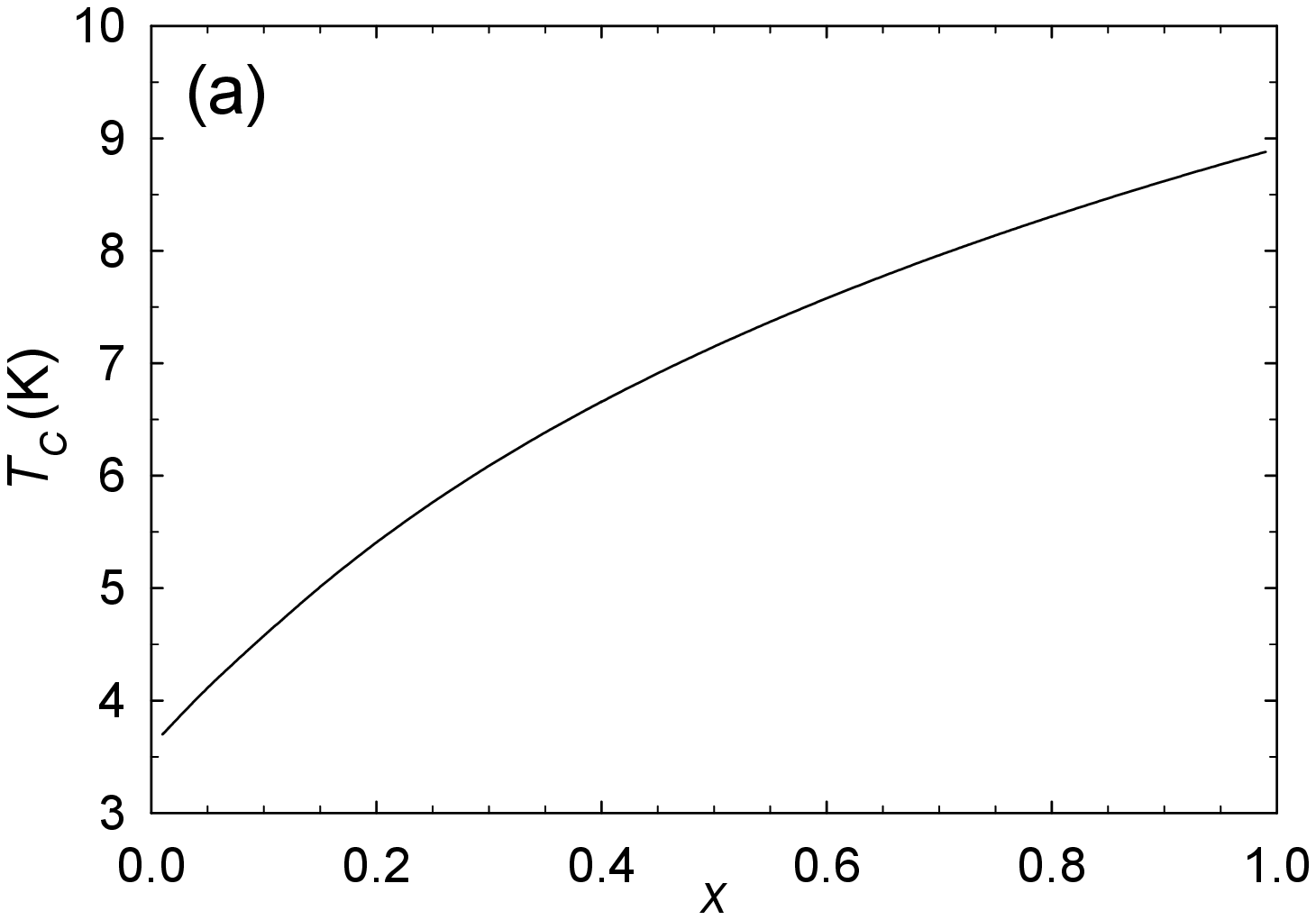}
\end{minipage}
\hspace{0.0 cm}
\begin{minipage}[b]{0.5\linewidth}
\includegraphics[width=2.9in]{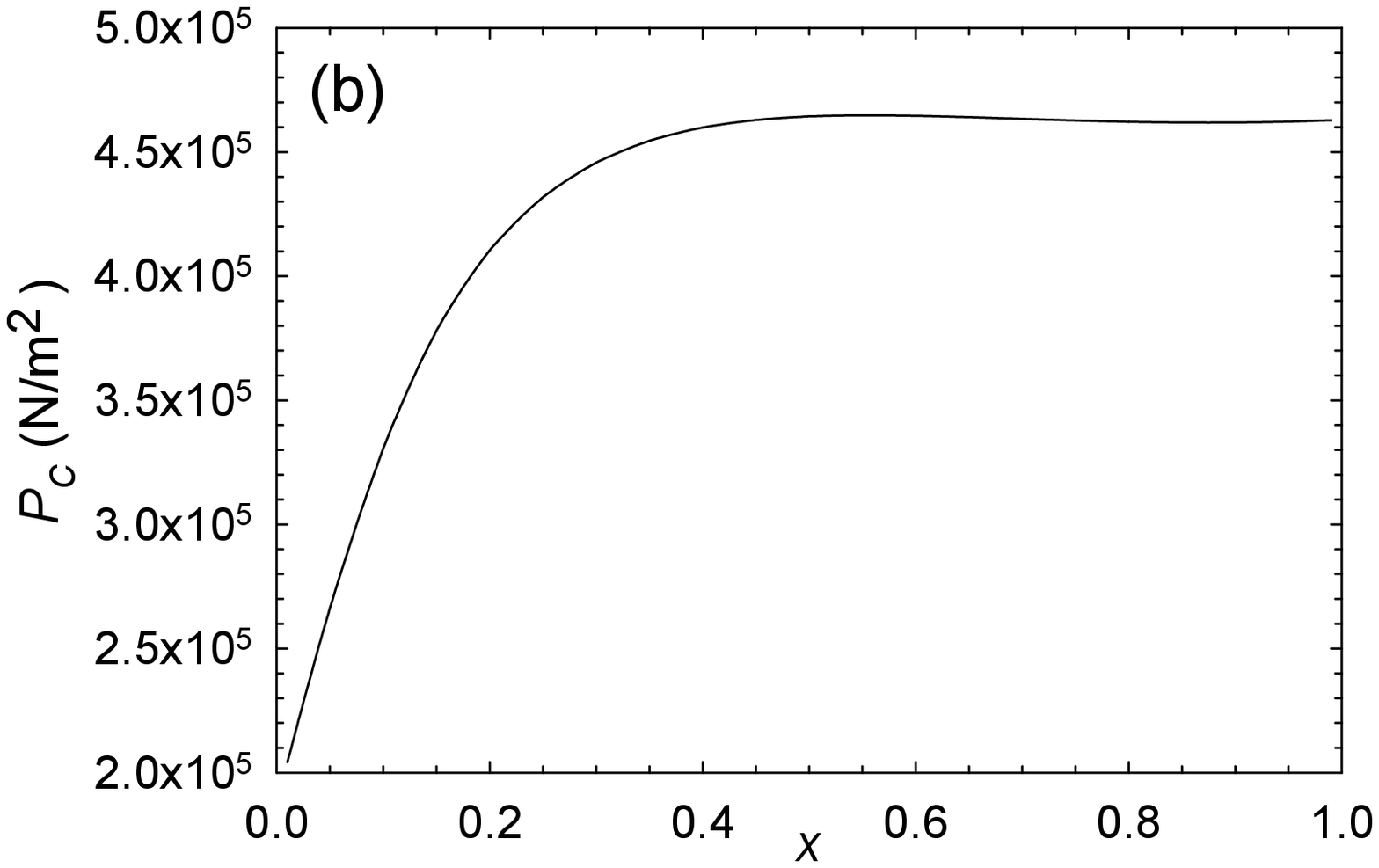}
\end{minipage}
\begin{center}
\includegraphics[width=2.9in]{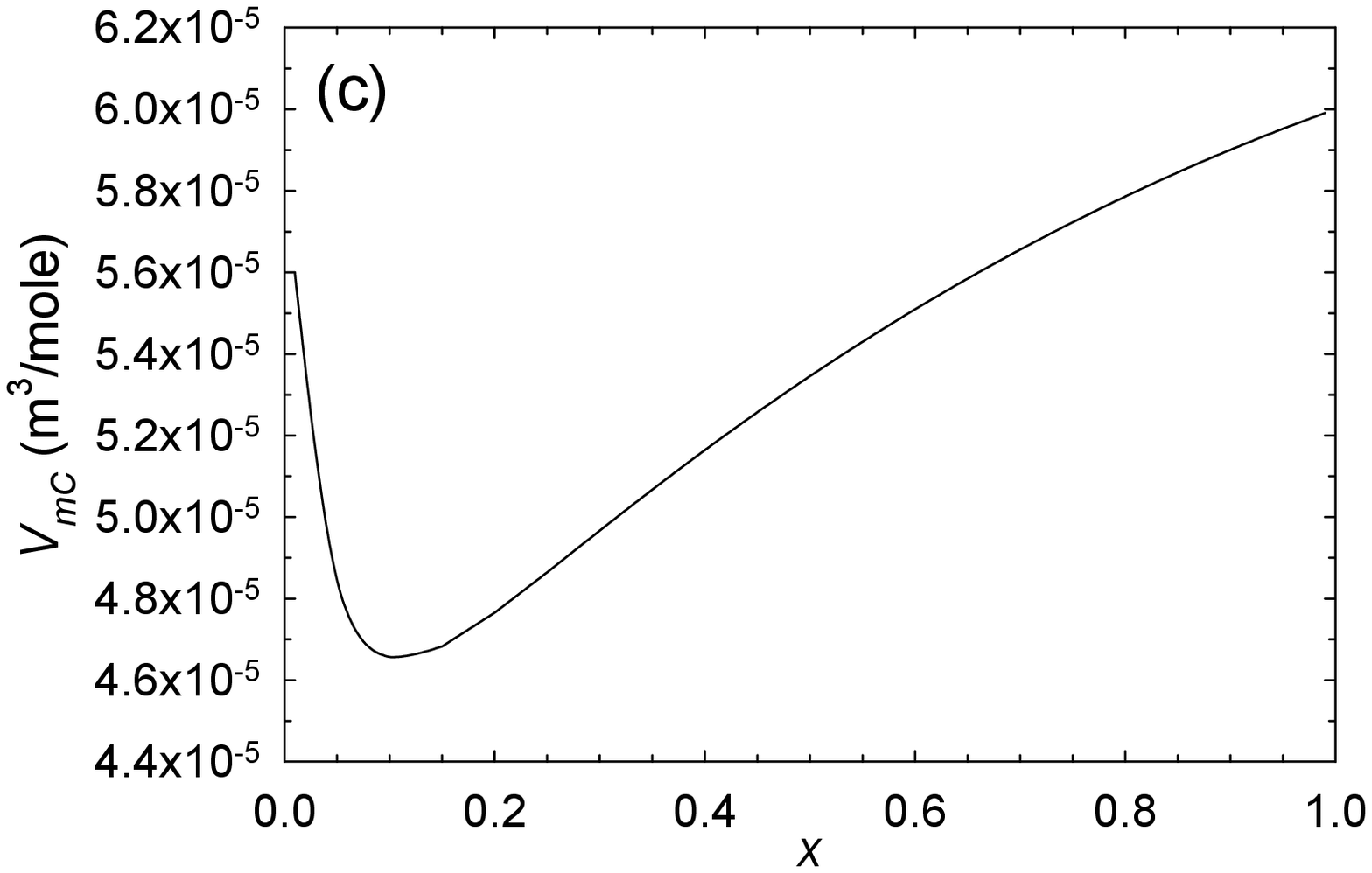}
\end{center}
\caption{The critical parameters $(T_c,P_c,V_{mc})$ as functions of $x$. On crossing the critical curve, $\det(\textbf{g})$ switches sign, with the thermodynamics stable above the critical curve and unstable below it. $R$ diverges to negative infinity on the critical curve.}
\label{Figure4}
\end{figure}

The critical curve is shown in Figure \ref{Figure4}. The $\det(\textbf{g})$ switches signs upon crossing this critical curve, with the thermodynamics stable above the curve and unstable below it, according to Eqs. (\ref{Stability1}), (\ref{Stability2}), and (\ref{Stability3}). $R$ diverges to negative infinity on the critical curve.

\par
The theoretical expectations on approaching the critical curve are well-known. A general physical argument was given by Widom \cite{Widom1974} in the context of the hyperscaling critical exponent relation
\begin{equation}d\nu=2-\alpha,\end{equation}

\noindent where $d$ is the spatial dimension ($d=3$ here), $\alpha$ is the heat capacity critical exponent ($\alpha=0$ here), and $\nu$ is the correlation length critical exponent. On approaching the critical curve from above along a critical isochore with fixed $x$, fixed $V_m=V_{mc}$, and decreasing $T$, the correlation length is expected to diverge according to:
\begin{equation}\xi^d\propto\left(\frac{T-T_c}{T_c}\right)^{\alpha-2}.\end{equation}

\noindent 
A key finding in previous research is that $|R|\propto \xi^d$ near a critical point \cite{Ruppeiner1995, Ruppeiner1979, Johnston2003, Ruppeiner2015},\footnote{The constant of proportionality between $R$ and $\xi^3$ calculated for both fluid and magnetic systems in distinct spatial dimensions appears to be $-1/2$: $-\frac{1}{2}R=\xi^d$; see \cite{Ruppeiner2015} for brief review.} and so we expect
\begin{equation}|R|\propto\left(\frac{T-T_c}{T_c}\right)^{-2}\label{PowerLaw}\end{equation}

\noindent along a critical isochore.

\par
An interesting contrast to binary vdW is offered by pure fluid vdW, where $R$ is computed with 2D thermodynamic geometry. It was shown \cite{Ruppeiner1995} that the asymptotic divergence of $R$ for the pure vdW fluid, with the thermal ideal gas contribution as in Eq. (\ref{IdealGas}), along the critical isochore is
\begin{equation}R=-b\left(\frac{T-T_c}{T_c}\right)^{-2},\label{RMPEquation}\end{equation}

\noindent independent of the vdW constant $a$. Here, $b=b_{11}$ if $x=0$ and $b=b_{22}$ if $x=1$.

\par
Physically, $\xi^3$ should be the same for a binary fluid in the limits $x\to 0$ or $x\to 1$ as for the corresponding pure fluids. Since $R$ in the critical region is a measure of $\xi^3$, we might then physically expect the limiting $R$ 3D's to be at least approximately equal to the corresponding $R$ 2D's. Remarkably, this limiting correspondence holds very well, even though the 2D and 3D formulas for $R$ are quite different.

\par
\begin{figure}
\centering
\includegraphics[width=4in]{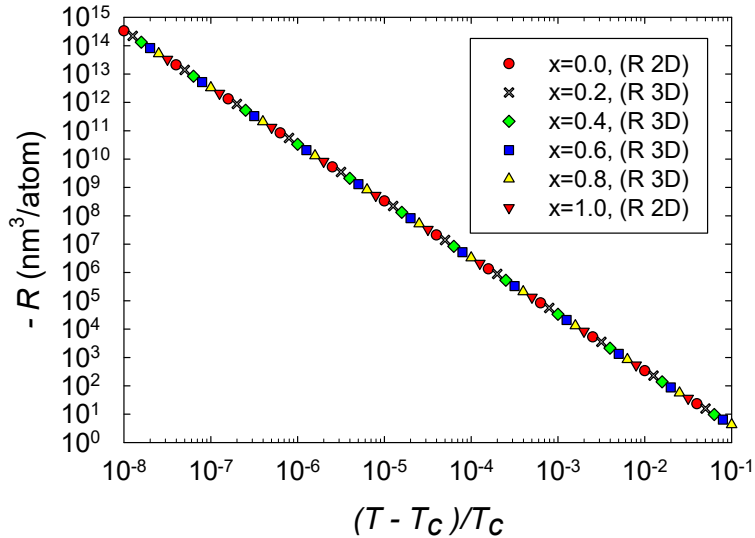}
\caption{$-R$ computed with the $R$ 3D formula versus the reduced temperature for four critical isochores in the critical region. Points all fall nearly on the same straight line with slope $-2$. The values of $R$ are all negative. Also shown is $-R$ computed with the $R$ 2D formula for the two pure fluids $x=0$ and $1$ along the critical isochores. Somewhat remarkably, these points also fall on the common line.}
\label{Figure5}
\end{figure}

We computed $R$ along four critical isochores in 3D thermodynamic geometry, $x=0.2, 0.4, 0.6$, and $0.8$. Results near the critical point are shown in Figure \ref{Figure5}. The four curves clearly overlap each other very closely, and show very little dependence on $x$. The asymptotic $R$ is always negative, in accord with expectations for effectively attractive interactions. The critical curves have asymptotic power law divergences, in accord with Eq. (\ref{PowerLaw}). The critical exponents of the four curves are all within better than $0.3\%$ of the expected value $2$, with the power law fits over the full range of the data in Fig. \ref{Figure5}. Erdem \cite{Erdem2018} got similar 3D power law behavior in the Ising antiferromagnet. Note as well that the smallest values of $|R|$ in Fig. \ref{Figure5} are about $4$ nm$^3$, so values near the particle level are on the asymptotic power law line. Such a large span of the critical regime was seen also in magnetic systems \cite{Ruppeiner2015}.

\par
Also shown in Fig. \ref{Figure5} is $-R$ for the pure fluids $x=0$ and $1$ calculated with the 2D thermodynamic geometry. Asymptotically, we expect these curves to obey Eq. (\ref{RMPEquation}), and we found that they do. The $R$ 2D curves are in excellent agreement with the corresponding $R$ 3D curves.

\par
Next, we consider letting the binary fluid expand into the binary ideal gas state, which has the dimensionless quantity \cite{Ruppeiner1990}

\begin{equation}R \times (\rho_1+\rho_2)=\frac{1}{2}\label{Ideal}.\end{equation}

\begin{figure}
\centering
\includegraphics[width=4in]{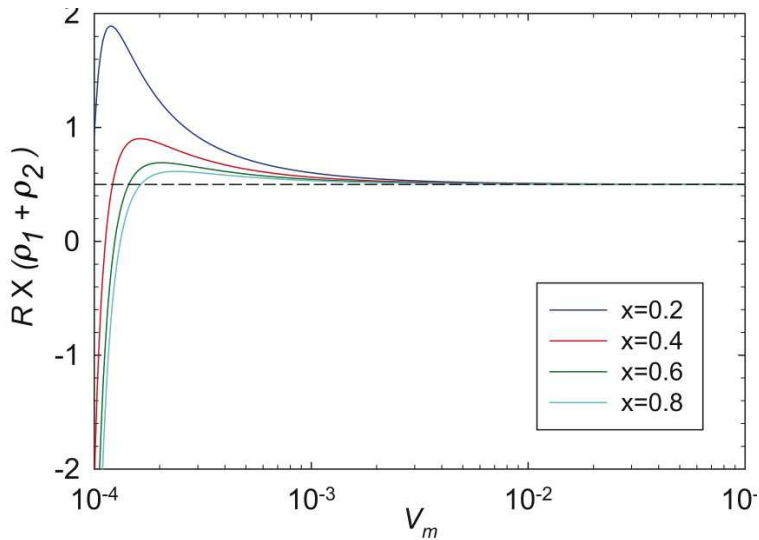}
\caption{The approach to the ideal gas limit. For each fixed $x$, we proceed at constant $T=T_c$ increasing $V_m$. This leads to the ideal gas where we expect $R\times(\rho_1+\rho_2) = 1/2$.}
\label{Figure6}
\end{figure}

\noindent We approached the binary ideal gas limit for four fixed values of $x$ by starting at $T_c$ for each $x$, and increasing $V_m$ at constant $T=T_c$. Results of this expansion are shown in Figure \ref{Figure6}. Clearly, for each $x$, $R\times(\rho_1+\rho_2)\to1/2$, in accord with Eq. (\ref{Ideal}). 

\section{Conclusion}

We have made a start on calculating the full three-dimensional thermodynamic Ricci curvature scalar $R$ in an interacting binary fluid system. Our main finding is that the physical interpretation of $R$ for the pure fluid extends very naturally to the binary fluid. The emerging physical picture for the thermodynamic $R$ is thus quite robust. We calculated $R$ for two scenarios involving the van der Waals model for the binary fluid. The first had exclusively repulsive interactions, and the second added attractive interactions, and critical phenomena.

\par
Let us place our results in the context of the expectations in Section 2. a) In the binary ideal gas limit, we expect $R \times (\rho_1+\rho_2)=1/2$ from an earlier exact calculation. Figure \ref{Figure6} clearly shows this known limit. This limit is also shown at the origin of Figure \ref{Figure2}. Both these figures (in the limit) have $|R|$ smaller than the low $|R|$ limit, and so the results here are not as strong as those in the critical regime, where $|R|$ is much larger. b) Near critical points, $|R|$ was found to diverge with the expected critical exponent of $2$, as shown in Figure \ref{Figure5}. In the limits as the binary fluid goes to the pure fluid, we found excellent concordance with the pure fluid results calculated with the 2D thermodynamic geometry. These critical point results are the strongest that we present in this paper. c) Near critical points, $R$ was found to be negative, as shown in Figure \ref{Figure5}. This negative sign is expected when interactions are effectively attractive. d) Evaluating $R$ along coexistence curves was beyond the scope of our paper. e) Our regimes of positive $R$ were all more or less expected, with values all under the low $|R|$ limit. A more systematic search of the thermodynamic phase space for more interesting cases (above the low $|R|$ limit) was beyond the scope of our paper. f) We found that the repulsive vdW was negative at large densities, in accord with the anomaly found by Bra\'nka, et al., for the hard sphere \cite{Branka2018}; see Figure \ref{Figure2}. That this anomaly translates from the 2D to the 3D thermodynamic geometry is interesting.


\newpage

\begin{thebibliography}{9}

\bibitem{Ruppeiner1995} G. Ruppeiner, ``Riemannian geometry in thermodynamic fluctuation theory," Rev. Mod. Phys. \textbf{67}, 605 (1995); Erratum \textbf{68}, 313 (1996).

\bibitem{Carroll2004} S. M. Carroll, {\it Spacetime and Geometry}, Addison Wesley, San Franciso (2004).

\bibitem{Sahay2017a} A. Sahay, ``Restricted thermodynamic fluctuations and the Ruppeiner geometry of black holes,'' Phys. Rev. D \textbf{95}, 064002 (2017).

\bibitem{Ruppeiner1990} G. Ruppeiner and C. Davis, ``Thermodynamic curvature of the multicomponent ideal gas,’’ Phys. Rev. A {\bf 41}, 2200 (1990).

\bibitem{Kaviani1999} K. Kaviani and A. Dalafi-Rezaie, ``Pauli paramagnetic gas in the framework of Riemannian geometry,’’ Phys. Rev. E {\bf 60}, 3520 (1999).

\bibitem{Erdem2018} R. Erdem, “Antiferromagnetic Ising model in the framework of Riemannian geometry,” Acta  Phys. Pol. B {\bf 49}, 1823 (2018).

\bibitem{Ginoza1993} M. Ginoza, ``Riemannian geometry of equilibrium thermodynamics in a liquid mixture,'' Reports on Mathematical Physics {\bf 32}, 167 (1993).

\bibitem{Arenas2020} J. Jaramillo-Gutiérrez, J. L. López-Picón, and J. Torres-Arenas, ``Thermodynamic geometry for binary mixtures: A constrained approach,’’ Journal of Molecular Liquids, preprint available online 8 September 2020, 114213.

\bibitem{Ruppeiner2010} G. Ruppeiner, ``Thermodynamic curvature measures interactions,'' G. Ruppeiner, American Journal of Physics {\bf 78}, 1170 (2010).

\bibitem{Ruppeiner2012a} G. Ruppeiner, A. Sahay, T. Sarkar, and G. Sengupta, ``Thermodynamic geometry, phase transitions, and the Widom line," Phys. Rev. E \textbf{86}, 052103 (2012).

\bibitem{Ruppeiner2012b} G. Ruppeiner, ``Thermodynamic curvature from the critical point to the triple point," Phys. Rev. E \textbf{86}, 021130 (2012).

\bibitem{May2013} H.-O. May, P. Mausbach, and G. Ruppeiner, ``Thermodynamic curvature for attractive and repulsive intermolecular forces," Phys. Rev. E \textbf{88}, 032123 (2013).

\bibitem{Branka2018} A. C. Bra\'nka, S. Pieprzyk, and D. M. Heyes, ``Thermodynamic curvature of soft-sphere fluids and solids," Phys. Rev. E \textbf{97}, 022119 (2018).

\bibitem{Callen1985} H. B. Callen, {\it Thermodynamics and an Introduction to Thermostatistics}, John Wiley \& Sons, New York (1985).

\bibitem{Landau1980} L. D. Landau and E. M. Lifshitz, {\it Statistical Physics} (Elsevier, New York, 1980).

\bibitem{Andresen1984} B. Andresen, P. Salamon, and R. S. Berry, ``Thermodynamics in finite time,’’ Phys. Today {\bf 37}, No. 9, 62 (1984).

\bibitem{Weinberg1972} S. Weinberg, {\em Gravitation and Cosmology}, Wiley, New York (1972).

\bibitem{Eves1966} H. Eves, {\it Elementary Matrix Theory}, Dover, New York (1966).

\bibitem{Scott1980} P. van Konynenburg and R. Scott, ``Critical lines and phase equilibria in binary van der Waals mixtures,'' Philosophical Transactions: A. Mathematical and Physical Sciences {\bf 298}, 495 (1980).

\bibitem{Widom1974} B. Widom, ``The critical point and scaling theory,’’ Physica {\bf 73}, 107 (1974).

\bibitem{Ruppeiner1979} G. Ruppeiner, ``Thermodynamics: A Riemannian geometric model,’’ Phys. Rev. A {\bf 20}, 1608 (1979).

\bibitem{Johnston2003} D. A. Johnston, W. Janke, and R. Kenna, ``Information geometry, one, two, three (and four),'' Acta Phys. Pol. B \textbf{34}, 4923 (2003).

\bibitem{Ruppeiner2015} G. Ruppeiner and S. Bellucci, ``Thermodynamic curvature for a two-parameter spin model with frustration,’’ Phys. Rev. E {\bf 91}, 012116 (2015).

\end{thebibliography}
\end{document}